\documentclass[aps,prl,showpacs,twocolumn,superscriptaddress]{revtex4}

\usepackage{epsfig}
\usepackage{graphicx}
\usepackage{amsmath}
\usepackage{amssymb}
\usepackage{amsfonts}
\usepackage{dcolumn}
\usepackage{color}

\begin{document}
\title{Dynamics of two atoms undergoing light-assisted collisions in an optical microtrap}

 \author{P. Sompet}
 \affiliation{Jack Dodd Center for Quantum Technology, Department of Physics, \\ University of Otago, New Zealand.}
\affiliation{Department of Physics and Materials Science, Faculty of Science, Chiang Mai University, Chiang Mai, Thailand 50200.}
 \author{A. V. Carpentier}
 \affiliation{Jack Dodd Center for Quantum Technology, Department of Physics, \\ University of Otago, New Zealand.}
 \author{Y. H. Fung}
 \affiliation{Jack Dodd Center for Quantum Technology, Department of Physics, \\ University of Otago, New Zealand.}
 \author{M. McGovern}
 \affiliation{Jack Dodd Center for Quantum Technology, Department of Physics, \\ University of Otago, New Zealand.}
 \author{M. F. Andersen}
 \affiliation{Jack Dodd Center for Quantum Technology, Department of Physics, \\ University of Otago, New Zealand.}

\begin{abstract}
We study the dynamics of atoms in optical traps when exposed to laser cooling light that induces light-assisted collisions. We experimentally prepare individual atom pairs and observe their evolution. Due to the simplicity of the system (just two atoms in a microtrap) we can directly simulate the pair's dynamics, thereby revealing detailed insight into it. We find that often only one of the collision partners gets expelled, similar to when using blue detuned light for inducing the collisions. This enhances schemes for using light-assisted collisions to prepare individual atoms and affects other applications as well.
\end{abstract}


\maketitle

Optically trapped cold atoms provide an exciting platform for studying the change in atom-atom interactions due to absorption or emission of light. The advantage of such a system is that the atoms are isolated from their surroundings and basic physical effects in photo-chemistry can be studied without interference from processes induced by the environment. Recently this has led to controlled formation of ultra-cold molecular gases \cite{Homo, Hetero} and in the past, photo-association spectroscopy has provided detailed knowledge about atom-atom interaction potentials \cite{Heinzen2, Lett}.

Typically, experiments on photo-association or light-assisted collisions of cold atoms are conducted on large samples. Studying microscopic processes at the individual event level reveals information hidden in ensemble averages of larger samples \cite{Toschek,Chu}. Pioneering work in studying individual light-assisted collisions was conducted using small samples of atoms in a high gradient Magneto-Optical Trap (MOT) \cite{Meshede1, Meshede2}. Those studies observed that up to $10\%$ of loss events manifested themselves as just one of the collision partners being lost from the MOT.

Of particular interest to many modern experiments in atomic physics are light-assisted collisions or photo-association events induced by near-resonant light. This phenomenon is of fundamental interest \cite{Browaeys} and plays a crucial role in modern laser cooling experiments. In addition, light-assisted collisions have been employed to isolate individual atoms in optical microtraps \cite{DePueMT, Schlosser, Schlosser2, Weiss, Andersen0}, to redistribute atoms loaded into an array of traps \cite{Meshede3}, and to perform parity number measurement of atoms in optical lattices \cite{Bloch1, Greiner}. Isolation experiments that used blue-detuned light to induce collisions have achieved high efficiency, whereas experiments using red detuned light have reported efficiencies of about 50\%.  
For several of these applications it is assumed that both atoms of the pair are lost from the optical microtrap when they undergo light-assisted collisions. 

\begin{figure}
\epsfig{figure=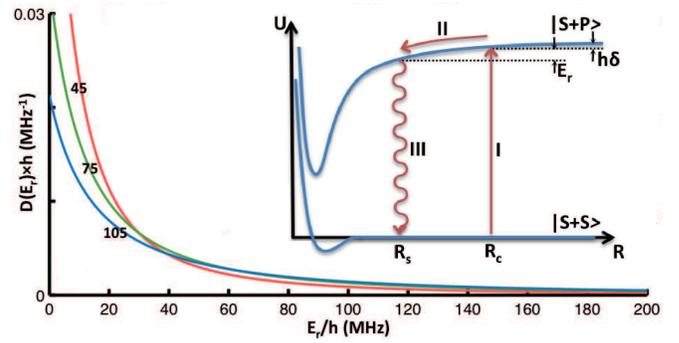, width=1\columnwidth}
\caption{
(Color online)
Inset: Molecular interaction potentials as a function of the inter-nuclear separation $R$.
Main graph: Probability density of the released energy for a relative collision speed of $v \left( R_c \right)= 0.2$ m/s for three detunings $\delta$ ($45, 75,$ and $105$ MHz).
}
\label{fig1}
\end{figure}

Here we revisit the ejection of atoms from a far off resonance optical trap due to light-assisted collisions induced by red-detuned laser cooling light. 
We implement the idealized collision experiment in which only two atoms are trapped so we can observe individual atom loss events. 
A numerical model of the complex dynamics of the expelling process agrees well with our experiment. 
We find that the light-assisted collisions can lead to just one of the collision partners being lost besides the pair loss often assumed. Generally both channels are present; but which is more likely depends on the dynamics of the atoms in the trap under the influence of laser cooling. The existence of collisional single atom ejection allows us to exceed the $50\%$ isolation efficiency of individual atoms previously reported when using red-detuned light-assisted collisions \cite{DePueMT, Schlosser, Schlosser2, Weiss}. The onset of other loss mechanisms still limits our single atom loading efficiency to $63\%$.

The inelastic collision process between two cold atoms induced by red-detuned light can be understood using a semiclassical model (illustrated in the inset of Fig. \ref{fig1}) \cite{GP,JV}.
The $\left| S+S \right\rangle $ asymptote represents two ground state atoms and $\left| S+P \right\rangle $ represents one ground and one excited state atom. As two atoms approach in a collision, the transition frequency changes, and at an inter-nuclear separation of $R=R_c$ (the Condon point) it is resonant with the laser field. The atoms may absorb a photon and get transferred to the excited molecular state (step I in the inset of Fig. \ref{fig1}). Here they attract and accelerate towards each other (II) until a spontaneous emission to the ground state occurs (III). The process releases an energy $E_r$ given by the difference between the excited state interaction energy at $R_c$ and at the inter-nuclear separation at which the spontaneous emission occurred ($R_s$). The main graph in Figure \ref{fig1} shows examples of the probability densities of the released energy $D(E_r)$ for different collision parameters calculated as described in \cite{sub}. In all cases, due to the shallow nature of the excited molecular state at $R_c$, it is probable that the atom pair decays to the ground state before a significant relative acceleration has occurred.

\begin{figure}
\epsfig{figure=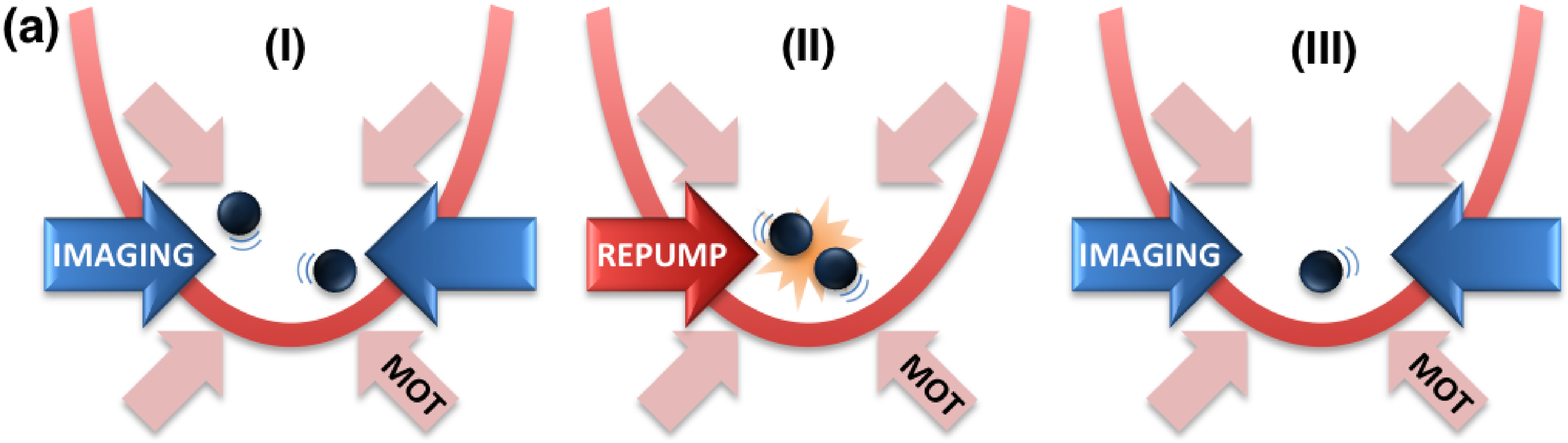, width=1\columnwidth}
\epsfig{figure=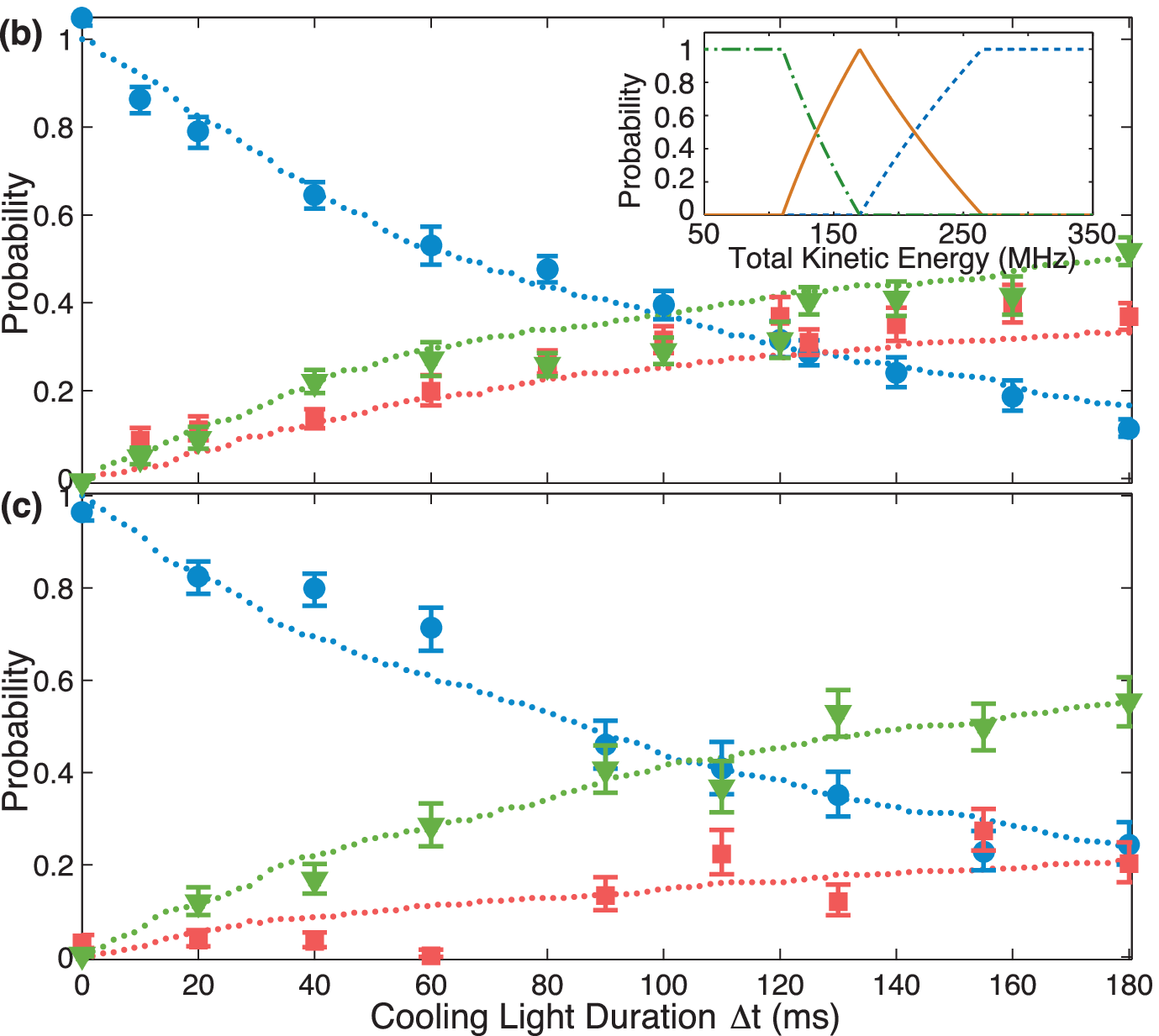, width=1\columnwidth}
\caption{
(Color online)
(a): Experimental sequence.
(b) and (c) show atom pairs evolution as a function of $\Delta t$ for MOT beam intensities of $11$ W/m$^2$ in (b) and $19$ W/m$^2$ in (c).
The blue circles indicate the survival probability of the pair. The red squares and the green triangles
show the probabilities of obtaining one and zero atoms respectively after the cooling pulse.
Error bars represent a statistical confidence of $68.3\%$.
Dotted lines are the simulation.
Inset: Probabilities for losing zero (green dashed-dotted line), one (orange line), and two atoms (blue dashed line) as a result of a light-assisted collision. 
}
\label{fig2}
\end{figure}

Figure \ref{fig2}(a) illustrates the procedure for our experimental study of collisions between the atoms of individual pairs.
We first prepare a few $^{85}$Rb atoms in a $U_0=h\times85$ MHz deep optical microtrap using the method described in \cite{Andersen0}. The microtrap laser wavelength is $828$ nm, and the MOT quadrupole magnetic field used during the initial loading stage is extinguished throughout the procedure illustrated.
In frame I, we measure the atom number using the fluorescence detection technique of \cite{Andersen2} and select realizations with a pair present.
In II, we induce collisions between the atoms and laser-cool them using a repump beam and the six cooling beams also used for the MOT. In combination we denote these beams the cooling light. An atom that after an inelastic collision has an energy larger than $U_0$ will escape. Collisions predominantly occur at the microtrap center where the atomic density is highest. The large detuning of the microtrap prevents it from inducing light-assisted collisions.
The repump beam's estimated intensity is $6.25$ W/m$^2$ at the position of the atoms and is, unless otherwise stated, red detuned by $\delta_c=45$ MHz from the $D1$ $F=2$ to $F'=3$ transition at the center of the trap. 
The cooling beams are red detuned by $16$ MHz from the $D2$ $F=3$ to $F'=4$ for a free atom. 
At its center the trap light shifts the atomic transitions such that the cooling beams are $4$ MHz red detuned from the $F=3$ to $F'=3$ transition averaged across all magnetic sub-levels. The cooling beams thereby provide effective optical pumping into the F=2 ground state. As the repump light is shifted away from resonance most colliding pairs will be in this state. The probability for a light-assisted collision depends on the detuning of the light that induces it \cite{wiener}. For the relatively low intensities we use the probability increase closer to resonance. Most light-assisted collisions are therefore induced by the repump light due to its relatively close proximity to a transition from the $F=2$ ground state.
We vary the cooling light duration $\Delta t$ to explore the time evolution of the pair.
Finally, in frame III, we measure the number of atoms remaining.
By averaging over $\sim 180$ runs with a pair initially present, we obtain the probabilities for ending with zero, one, and two atoms.

The measurements shown in Fig. \ref{fig2}(b) and (c) display a nonzero probability of a single atom loss event when the red-detuned cooling light induces collisions. 
For light-assisted collisions to induce single atom loss, the two atoms must have different energies after the collision such that only one of them has enough energy to escape the trap. The difference in kinetic energy depends on the angle between the pair's center of mass velocity  $\mathbf{v}_\mathrm{CM}=\left(\mathbf{v}_1+\mathbf{v}_2 \right)/2$ and its relative velocity $\mathbf{v}_\mathrm{R}=\mathbf{v}_1-\mathbf{v}_2 $, as well as their magnitudes ($\mathbf{v}_1$ and $\mathbf{v}_2$ being the velocities of atoms 1 and 2).
The center of mass velocity is unaffected by the collision. Assuming that the direction of the relative velocity is uncorrelated with the center of mass velocity, one can compute the probability that one of the atoms will be lost while the other remains trapped. For a total kinetic energy of the pair after a collision given by $K_p=E_r+K_I$, with $K_I$ the initial kinetic energy, the probability for one atom to escape is  $P_1=1-\frac{\left| 2 \left| U\left(\mathbf{x} \right) \right| -K_p \right|}{2 m v_\mathrm{CM} \sqrt{\frac{K_p}{m}-v_\mathrm{CM}^2}}$ when $K_p\in \left[ E_\mathrm{CM} -2 v_\mathrm{CM} \sqrt{2 \left| U\left( \mathbf{x} \right) \right| m} ; E_\mathrm{CM} +2 v_\mathrm{CM} \sqrt{2 \left| U\left( \mathbf{x} \right) \right| m} \right]$ and $P_1=0$ elsewhere. $m$ is the mass of an atom, $U\left( \mathbf{x} \right)$ is the microtrap potential at the position $\mathbf{x}$ where the collision happens (defined using $U=0$ far from the trap), and $E_\mathrm{CM}=2 \left| U\left( \mathbf{x} \right) \right| + 2 m v_\mathrm{CM}^2$. The inset in Fig \ref{fig2}(b) shows $P_1$ for a collision occurring at the trap center with $v_\mathrm{CM}=20$ cm/s, which is a typical center of mass speed for atoms with a temperature of a few hundred micro Kelvin. Additionally, the inset shows the probability of not losing any atom ($P_0=1-P_1$ for $Kp<2 \left| U\left( \mathbf{x} \right) \right|$ and $0$ elsewhere) and the probability for losing both atoms ($P_2=1-P_1$ for $K_p>2\left| U\left( \mathbf{x} \right) \right|$ and $0$ elsewhere). We see that a high $E_r$ collision, which has low probability as seen in Fig. \ref{fig1}, leads to pair loss, while both atoms remain when a low energy is released. For a broad range of intermediate $E_r$s the nonzero center of mass speed makes it possible to only lose one atom.   
Between collisions that do not lead to loss, a fast laser cooling rate favors the removal of the energy released. The temperature of a colliding pair, which determines the typical $v_\mathrm{CM}$s in a collision, is then the laser cooling equilibrium temperature.
The laser cooling parameters therefore play a crucial role for the dynamics of the pair. This is observed in Fig. \ref{fig2}(b) and (c) where different cooling beam powers result in different probabilities for collisions leading to single atom or pair loss. 

\begin{figure}
 \epsfig{figure=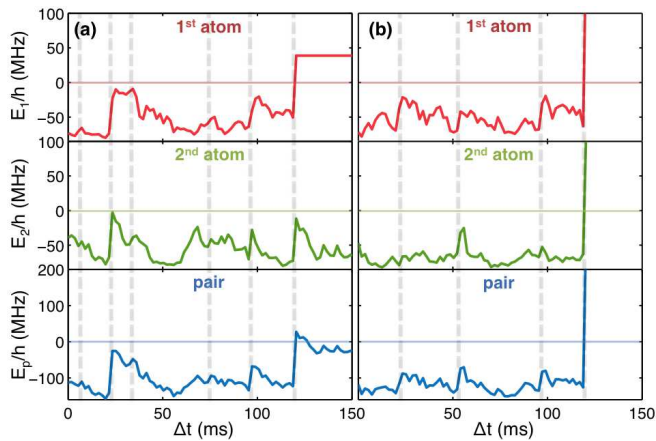, width=1\columnwidth}
\caption{
 (Color online) Evolution of the individual and the combined energy of a pair. (a): a run ending with single atom loss and (b): one ending with the pair loss. The dashed lines show when inelastic collisions occur.
 }
\label{fig3}
\end{figure}

To test the above explanation we perform a numerical simulation of the experiment. In it two atoms are initially randomly selected from the Maxwell-Boltzman distribution with the initial temperature of the pairs ($\sim280$ $\mu$K in our experiment). Between collisions their classical trajectories in the Gaussian potential are then computed. Laser cooling during their motion is simulated by a Doppler cooling model (for details see \cite{sub}). When the two atoms reach an inter-nuclear separation of $R=R_c$, they may undergo an inelastic collision with probability $P_e$. $P_e$ is determined by using the Landau-Zener formalism on a two level molecular model in the dressed state picture \cite{wiener, sub}. When an inelastic collision occurs we compute $E_r$ using a semiclassical model \cite{GP, JV}. The atoms interact due to their excited state molecular interaction potential, while their relative position is treated classically. During this motion they can spontaneously decay to the ground state and $E_r$ is found as the difference in interaction energy at $R_c$ and $R_s$ \cite{sub}. Finally, $E_r$ is transferred to the pair such that its center of mass momentum is conserved and the change in the individual atoms' momentum is along their inter-nuclear axis.

Figure \ref{fig3} shows the evolution of the individual energies of two atoms ($E_1$ and $E_2$) and their combined energy ($E_p=E_1+E_2$) in a simulation run leading to single atom loss (a) and another leading to pair loss (b). The gray dashed lines indicate when inelastic collisions occur. Most of these release a relatively low energy and no atoms are lost. The atoms generally share the released energy unevenly leading to single atom ejection unless a high energy is released in a single collision. In such cases both atoms are lost as can be seen in Fig. \ref{fig3}(b). The reduction of energy between collisions by laser cooling prevents collisions to effectively cease as the density drops at high energy.

The results of simulation are displayed alongside those of the experiment in Fig. \ref{fig2}(b) and (c), showing good agreement. The probability for single atom ejection decreases when we increase the cooling beam intensity. 
High cooling beam intensities generally provide more efficient cooling lowering the typical $E_p$ before a collision. This lowers the chance of single atom ejection by reducing the probability that the atoms share the energy unevenly and increasing the required $E_r$.

\begin{figure}
 \epsfig{figure=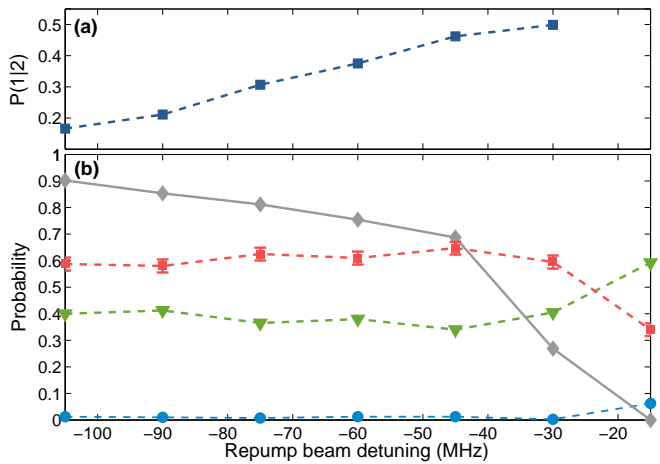, width=1\columnwidth}
\caption{
 (Color online)
 (a): Probability of collisional one atom loss $P\left(1|2 \right)$ as a function of repump beam detuning $\delta_c$.
 (b): Single atom loading probability versus $\delta_c$ (red squares).
Green triangles and blue circles show the probabilities of obtaining zero and two atoms respectively. 
The gray line is the single atom survival probability after a cooling light pulse with $\Delta t=1.5$ s. 
The lines are a guide to the eye. }
\label{fig4}
\end{figure}

In Fig. \ref{fig1} we observed that $D(E_r)$ depends on $\delta$. Large detunings favor large energy releases due to the larger gradient of the excited molecular state at smaller $R_c$. Since pair loss is caused by large $E_r$ collisions, the probability for collisional single atom loss should depend on the repump beam's detuning. We study this by measuring the pair's evolution in a similar manner to Fig. \ref{fig2} for a range of $\delta_c$. The inelastic collision rate has a strong dependence on $R_c$ which depends on $\delta_c$. To keep the collision rate similar for each $\delta_c$ we adjusted the repump beam power to keep a pair decay time of $\sim90$ ms, (the other parameters were kept as in Fig. \ref{fig2}(b)). We then determine the probability that a collisional loss event leads to only one of the atoms being lost ($P(1|2)$) as described in \cite{sub}, and show the result in 
Fig. \ref{fig4}(a). As expected, we observe that for large detunings both partners are typically lost. A repump beam close to resonance yields a short single atom lifetime ($\tau$) due to heating caused by radiation pressure. When pair decay is dominated by the finite $\tau$, the measurement of $P(1|2)$ becomes inaccurate, limiting our measurements to $\delta_c \le -30$ MHz.

Our demonstration of a non-zero $P\left( 1|2 \right)$ affects the applications of light-assisted collisions. Whereas it obscures parity measurements it may enhance the efficiency $P$ beyond $50 \%$ when light-assisted collisions are used for the isolation of individual atoms in optical microtraps \cite{DePueMT, Schlosser, Schlosser2, Weiss}. A high $P$ is important to applications where multiple traps have to be loaded each with one atom simultaneously \cite{Saffman}. 
To investigate $P$ we prepare about 30 atoms in the microtrap and expose them to cooling light to isolate an atom from the sample. If atoms were lost in pairs we would have a $50\%$ chance for ending with one atom depending on whether the initial number was even or odd \cite{DePueMT}. However, for infinite $\tau$ a nonzero $P\left(1|2 \right)$ gives rise to a probability for ending with exactly one atom that exceeds $50\%$ given by $P=\frac{1}{2-P\left(1|2 \right)}$ \cite{Andersen0}. A finite $\tau$ reduces $P$ as a prepared single atom may be lost before detection.

Figure \ref{fig4}(b) shows $P$ as a function of $\delta_c$ and the probabilities of ending with zero and two atoms. 
$P$ initially increases slightly as the magnitude of $\delta_c$ decreases agreeing with the trend of $P \left( 1|2 \right)$ in Fig. \ref{fig4}(a).
Close to resonance, the rise of $P$ is obstructed by the short $\tau$. 
As a measurement of $\tau$ the gray line in Fig. \ref{fig4}(b) shows the survival probability of a single atom exposed to cooling light for $1.5$ s ($SP=\exp \left(-\frac{1.5s}{\tau} \right)$). It is determined by preparing a single atom, exposing it to a cooling light pulse of duration $1.5$ s, and finally measuring the probability that the atom remains. For a wide range of parameters, $P$ exceeds $50\%$. To confirm this, $1000$ experimental runs for $\delta_c=-45$ MHz yields $P=63 \pm 1.6\%$. This is still less than what can be achieved using blue detuned light \cite{Andersen0}. $P=63 \%$ occurs as a compromise between $P \left( 1|2 \right)$ and $\tau$. 

To further investigate the dependence of $P\left(1|2 \right)$ and $P$ on experimental parameters, we varied the trap depth $U_0$ by changing the power of the microtrap beam. This changes the ratio between the single atom equilibrium temperature and $U_0$ (and thereby $\tau$) as well as the $E_r$ required for one or both atoms to be lost after an inelastic collision. We again observe that when $\tau$ is small then $P\left(1|2 \right)$ is large. Although these effects partly counteract we observe a monotonic increase in $P$ with $U_0$.

In future work it would be interesting to use a tight trap geometry that yields a high rate of inelastic collisions. This could provide a high temperature of the colliding pair without needing to compromise the efficiency of the laser cooling. It may result in a high $P \left( 1|2 \right)$ without compromising the single atom equilibrium temperature and thereby $\tau$. We expect that $P$ could be improved considerably under such conditions.

In summary, we have studied the dynamics of the expelling process of optically trapped atoms due to light-assisted collisions induced by our cooling lasers. We prepared individual atom pairs and studied their evolution in an optical microtrap when exposed to cooling light. We found that light-assisted collisions, for most parameters investigated, can cause loss of only one of the collision partners in addition to the pair loss observed before. This finding highlights the importance of studying microscopic processes at the individual event level as it allows us to discriminate between pair and single atom loss. Furthermore, experiments with just two atoms allow for numerical modeling of the process, giving detailed insight into it. The numerical simulation agrees surprisingly well with our experimental results, considering the simple two-state semiclassical model used to simulate the energy release in the light-assisted collisions.
Our findings may have important implications for applications of light-assisted collisions. It could affect the interpretation of parity measurements \cite{Bloch1, Greiner}. Moreover, we show that the loading efficiency of single atoms can exceed the $50\%$ limit found in similar experiments \cite{DePueMT, Weiss} and those using the collisional blockade variation \cite{Schlosser, Schlosser2}.
Finally, our demonstration of photo association of individually prepared pairs of atoms marks an initial step towards being able to assemble individual complex molecules atom by atom.


\emph{Acknowledgments.-} 
This work is supported by UORG.
P. Sompet is funded from DPST and acknowledges partial support from ThEP.

\end{document}